\documentclass[aps,twocolumn,superscriptaddress,prd]{revtex4-2}
\usepackage{amsmath}
\usepackage{amssymb}
\usepackage{latexsym}
\usepackage{amsfonts}
\usepackage{epsfig}
\usepackage{psfrag}
\usepackage{graphicx}
\usepackage{color}
\usepackage{hyperref}
\usepackage{bbold}
\usepackage{graphicx}

\usepackage{amsmath}

\begin{document}

 \title{Higher-harmonic generation in the driven Mott-Hubbard model}

\author{Friedemann Queisser}

\affiliation{Helmholtz-Zentrum Dresden-Rossendorf, 
Bautzner Landstra{\ss}e 400, 01328 Dresden, Germany,}

\affiliation{Institut f\"ur Theoretische Physik, 
Technische Universit\"at Dresden, 01062 Dresden, Germany,}

\email{f.queisser@hzdr.de}
\author{Ralf Sch\"utzhold}

\affiliation{Helmholtz-Zentrum Dresden-Rossendorf, 
Bautzner Landstra{\ss}e 400, 01328 Dresden, Germany,}

\affiliation{Institut f\"ur Theoretische Physik, 
Technische Universit\"at Dresden, 01062 Dresden, Germany,}

\date{\today}

\begin{abstract}
Using Floquet theory and the hierarchy of correlations, 
we study the non-equilibrium dynamics of the 
Mott insulator state in the Fermi-Hubbard model under the influence of 
a harmonically oscillating electric field representing the pump laser. 
After deriving the associated Floquet exponents, we consider 
higher-harmonic generation where the strongest signal is obtained if the 
driving frequency equals one third of the Mott gap. 
\end{abstract}

\maketitle

\section{Introduction}

Non-equilibrium dynamics of interacting quantum many-body systems are a 
rich and complex field displaying many fascinating phenomena -- 
especially for strongly interacting systems. 
As the {\em drosophila} of strongly interacting quantum many-body systems,
we consider the Fermi-Hubbard model.
The non-equilibrium dynamics can be generated by an external stimulus, 
such as a pump laser \cite{G09,L19,L09,K18}, which we model as a harmonically oscillating 
electric field -- facilitating a Floquet analysis \cite{T08,W19}. 

Even for a static electric field, the Fermi-Hubbard model displays 
interesting phenomena \cite{E11,A12,J15,E14}, such as doublon-holon pair creation 
(dielectric breakdown, see, e.g., \cite{E10,P14,E13,L12,L14,O10}) 
which can also be understood in a Floquet picture. 
For harmonically oscillating fields, one can employ Floquet engineering
\cite{W21,M18c}, for example in order to suppress dissipation \cite{V21} 
or to modify pre-thermalization dynamics \cite{H17,F14}.
In the following, we shall study higher-harmonic generation in the 
harmonically driven Fermi-Hubbard model \cite{M21,S18,M18,M18b}. 
To this end, we combine Floquet theory with the method of the hierarchy
of correlations \cite{Nav10}. 
Note that, in contrast to many other approaches, we do not apply the 
Floquet expansion directly to the quantum state \cite{S22}, but to the time-dependent
correlation functions. 








\section{Driven Fermi-Hubard model}

To analyze the dynamics of a prototypical strongly interacting quantum 
many-body system under periodic driving, we consider the fermionic 
Hubbard model coupled to an external electric field. 
The Hamiltonian governing the system is given by ($\hbar=1$)
\begin{align}\label{extHubb}
\hat{H}_\mathrm{ext}=-\frac{1}{Z}\sum_{\mu\nu s}T_{\mu\nu}\hat{c}_{\mu s}^\dagger\hat{c}^{}_{\nu s}
+U \sum_\mu \hat{n}_\mu^\uparrow \hat{n}_\mu^\downarrow
+ \sum_{\mu s}V_\mu(t)\hat{n}_{\mu s}\,.
\end{align}
Here, $V_\mu(t)=q\mathbf{x}_\mu \cdot \mathbf{E}(t)$ represents the 
time-dependent potential energy of an electron with charge $q$ at the 
position $\mathbf{x}_\mu$ of site $\mu$, where $\mathbf{E}(t)$ is 
the external electric field.
%
Assuming that the wavelength of the pump laser is much larger than all 
other relevant length scales, we neglect its spatial dependence and 
approximate it by a purely time-dependent external electric field 
$\mathbf{E}(t)$. 

The fermionic creation and annihilation operators at sites $\mu$ and $\nu$, 
with spin $s\in{\uparrow,\downarrow}$, are denoted as 
$\hat{c}^\dagger_{\mu s}$ and $\hat{c}_{\nu s}$, respectively.
The hopping matrix $T_{\mu\nu}$ takes the value of the tunneling rate $T$ 
for nearest neighbors, and zero otherwise. 
The coordination number $Z$ represents the number of nearest neighbors 
for a given lattice site.

To account for the potential term, we perform a Peierls substitution, 
leading to complex-valued entries for the hopping matrix. 
The time-dependent hopping matrix is then expressed as
\begin{align}
T_{\mu\nu}(t)&=T_{\mu\nu}
\exp\left\{-i\int_0^t dt'[V_\mu(t')-V_\nu(t')]\right\}
\label{peierls2}\,. 
\end{align}
In this way, we obtain the explicitly time-dependent Hamiltonian
\begin{align}\label{PeierlsHubb}
\hat{H}(t)
=
-\frac{1}{Z}\sum_{\mu\nu s}T_{\mu\nu}(t)
\hat{c}_{\mu s}^\dagger\hat{c}^{}_{\nu s}
+U \sum_\mu \hat{n}_\mu^\uparrow \hat{n}_\mu^\downarrow
\,.
\end{align}
Similar to a gauge transformation in electrodynamics, the scalar potential 
$V_\mu(t)$ is transferred to the hopping matrix $T_{\mu\nu}(t)$, whose 
phase is analogous to that given by the vector potential, while the 
on-site repulsion term $U$ remains unchanged. 

\subsection{The hierarchy of correlations}

To obtain approximate solutions, we utilize a hierarchical method 
suitable for systems with large coordination numbers $Z$. 
To do this, we separate the reduced density matrices into 
correlations between lattice sites and on-site density matrices. 
For example, the correlation between two sites is given by $\hat{\rho}^\mathrm{corr}_{\mu\nu}=\hat{\rho}_{\mu\nu}-\hat{\rho}_\mu\hat{\rho}_\nu$. Likewise, correlations among three sites can be expressed as $\hat{\rho}^\mathrm{corr}_{\mu\nu\lambda}=\hat{\rho}_{\mu\nu\lambda}-\hat{\rho}_\mu\hat{\rho}_\nu\hat{\rho}_\lambda-\hat{\rho}^\mathrm{corr}_{\mu\nu}\hat{\rho}_\lambda-\hat{\rho}^\mathrm{corr}_{\mu\lambda}\hat{\rho}_\nu-\hat{\rho}^\mathrm{corr}_{\nu\lambda}\hat{\rho}_\mu$, etc. 

The on-site density matrix $\hat{\rho}_\mu$
and the two-site correlators $\hat{\rho}_{\mu\nu}^\mathrm{corr}$
obey evolution equations, 
which can be schematically represented as follows \cite{Nav10}
\begin{align}
i\partial_t \hat{\rho}_\mu
&=
F_1(\hat{\rho}_\mu,
\hat{\rho}_{\mu\nu}^\mathrm{corr})=\mathcal{O}(1)
\label{onsite}
\\
i\partial_t \hat{\rho}_{\mu\nu}^\mathrm{corr}
&=
F_2(\hat{\rho}_\mu,\hat{\rho}_{\mu\nu}^\mathrm{corr},\hat{\rho}_{\mu\nu\lambda}^\mathrm{corr})=\mathcal{O}(1/Z)
\,.
\label{twosite}
\end{align}
Similar equations hold for higher-order correlations. 
The specific forms of the nonlinear functions $F_n$ are determined by the 
exact von Neumann equation for the density matrix of the Hubbard model. 
Analyzing the evolution equations for the correlators reveals \cite{Queiss14,Queiss23}
that the scaling hierarchy remains preserved over time when the initial 
values satisfy 
$\hat{\rho}_{\mu}=\mathcal{O}(1)$, 
$\hat{\rho}_{\mu\nu}^\mathrm{corr}=\mathcal{O}(1/Z)$, 
etc. 

As the two-site correlators are of order $1/Z$, we can approximate 
Equation~(\ref{onsite}) as 
$i\partial_t \hat{\rho}_\mu\approx F_1(\hat{\rho}_\mu,0)$. 
The zeroth-order solution $\hat{\rho}_\mu^0$ determines the mean field background of the system and will be discussed in the subsequent section.
Similarly, since the three-point correlators scale with $1/Z^2$, 
we can approximate Equation~(\ref{twosite}) as 
$i\partial_t \hat{\rho}_{\mu\nu}^\mathrm{corr}\approx 
F_2(\hat{\rho}_\mu,\hat{\rho}_{\mu\nu}^\mathrm{corr},0)$.
Introducing the following operators proves to be advantageous when 
analyzing the evolution equation
\begin{align}
\hat c_{\mu s I}=\hat c_{\mu s}\hat n_{\mu\bar s}^I=
\left\{
\begin{array}{ccc}
 \hat c_{\mu s}(1-\hat n_{\mu\bar s}) & {\rm for} & I=0 
 \\ 
 \hat c_{\mu s}\hat n_{\mu\bar s} & {\rm for} & I=1
\end{array}
\right.
\,.
\end{align}
Here $\bar{s}$ denotes the opposite spin to $s$.
The dynamics of the two-site correlation functions to first order 
$\mathcal{O}(1/Z)$ takes then the form
\begin{align}
\label{corr-evolution}
i\partial_t
\langle\hat c^\dagger_{\mu s I}\hat c_{\nu s J}\rangle^{\rm corr}
=
\frac1Z\sum_{\lambda L} T_{\mu\lambda}(t)
\langle\hat n_{\mu\bar s}^I\rangle^0
\langle\hat c^\dagger_{\lambda s L}\hat c_{\nu s J}\rangle^{\rm corr}
\nonumber\\
-
\frac1Z\sum_{\lambda L} T_{\lambda\nu}(t)
\langle\hat n_{\nu\bar s}^J\rangle^0
\langle\hat c^\dagger_{\mu s I}\hat c_{\lambda s L}\rangle^{\rm corr}
\nonumber\\
+
U(J-I)
\langle\hat c^\dagger_{\mu s I}\hat c_{\nu s J}\rangle^{\rm corr}
\nonumber\\
+\frac{T_{\mu\nu}(t)}{Z}
\left(
\langle\hat n_{\mu\bar s}^I\rangle^0
\langle\hat n_{\nu s}^1\hat n_{\nu\bar s}^J\rangle^0
-
\langle\hat n_{\nu\bar s}^J\rangle^0
\langle\hat n_{\mu s}^1\hat n_{\mu\bar s}^I\rangle^0
\right) 
\,,
\end{align}
where the expectation values $\langle\hat X_\mu\rangle^0$ are taken 
with the zeroth-order solution 
$\langle\hat X_\mu\rangle^0={\rm Tr}\{\hat X_\mu\hat{\rho}_\mu^0\}$.
For a given mean-field background  $\hat{\rho}_\mu^0$, 
these differential equations describe 
the free dynamics of the quasi-particle excitations. 
In the absence of an electric field, Equation~(\ref{corr-evolution}) 
yields the dispersion of the quasi-particle excitations. 

While one approach to studying the equations of motion involves 
factorizing them {\cite{Nav14}, leading to an effective Dirac equation for the 
quasi-particle operators \cite{QS23}, in this case, we opt for a direct 
analysis focusing on the dynamics of the correlations.

\section{Floquet analysis}

In order to solve Equation~(\ref{corr-evolution}), we have so specify the 
mean-field background  $\hat{\rho}_\mu^0$.
In the strong-coupling regime $U\gg T$, the Mott insulator state has one 
particle per site (neglecting virtual hopping corrections with 
probabilities of order $T^2/U^2$), which merely leaves the spin structure 
to be determined. 

In the strong-coupling limit $U\to\infty$, any finite temperature 
would be below $U$ (i.e., the Mott gap), but above $T^2/U$ 
(i.e., the energy scale of spin fluctuations). 
Thus, such a temperature is sufficiently low to prevent the generation 
of doublon-holon pairs, but it would have the tendency to disrupt the 
spin order within the system. 
Consequently, we choose the following ansatz for a spin-disordered 
background 
\begin{align}\label{disorder}
\hat\rho_\mu\approx\hat\rho_\mu^0
=
\frac{|{\uparrow}\rangle_\mu\langle{\uparrow}|+
|{\downarrow}\rangle_\mu\langle{\downarrow}|}{2}
\,.
\end{align}
By using Equation~(\ref{onsite}), it can be verified that this state 
remains inert to lowest order $\mathcal{O}(1)$. 
Only when considering the back-reaction of the correlators on the mean 
field, an amplitude for the generation of doublon-holon pairs arises 
at next order $\mathcal{O}(1/Z)$.

Within the leading order of the hierarchical expansion, the modes of 
the quasi-particle excitations decouple. 
Specifically, after performing a Fourier transform of 
Equation~(\ref{corr-evolution}), we observe that each mode evolves 
according to
\begin{align}\label{firstorder}
i\partial_t f^{IJ}_{\mathbf{k}s}&=(J-I)Uf^{IJ}_{\mathbf{k}s}\nonumber\\
&+\frac{T_\mathbf{k}(t)}{2}\sum_L\left(f^{LJ}_{\mathbf{k}s}-f^{IL}_{\mathbf{k}s}\right)
+(I-J)\frac{T_\mathbf{k}(t)}{4}\,.
\end{align}
As explained above, the electrical field is encoded in the Fourier 
components of the hopping matrix, which adopts the usual minimal 
coupling form $T_\mathbf{k}(t)=T_{\mathbf{k}-q\mathbf{A}(t)}$ 
with $\partial_t \mathbf{A}(t)=-\mathbf{E}(t)$. 
For simplicity, we assume a periodic driving field of the form 
$\mathbf{E}=\mathbf{E}_0\cos(\omega t)$, i.e., in linear polarization, 
where 
$\omega$ is the driving frequency.

As the next step, we Fourier decompose the time dependence of 
$T_\mathbf{k}(t)$
\begin{align}
\label{channels}
T_\mathbf{k}(t)=T_\mathbf{k}^\perp
+\sum_n T_{\mathbf{k},n}^\|e^{-in\omega t}
\,,
\end{align}
where $T_\mathbf{k}^\perp$ contains the contribution perpendicular to 
the electric field (which is thus independent of time) while the 
$T_{\mathbf{k},n}^\|$ encode the time-dependence induced by the 
electric field (i.e., the Floquet channels). 

E.g., for a square lattice with the lattice spacing $\ell$, 
where the electric field $\mathbf{E}_0$ is aligned with one of the 
principal axes, these expansion coefficients $T_{\mathbf{k},n}^\|$ 
can be expressed in terms of Bessel functions of the first kind. 
For even values of $n$, the coefficients are real, taking the form 
$T_{\mathbf{k},n}^\|=
(2T/Z)\cos(k^\parallel\ell)\mathcal{J}_n(qE_0\ell/\omega)$. 
%
On the other hand, for odd values of $n$, the coefficients are 
purely imaginary, given by 
$T_{\mathbf{k},n}^\|=-
i(2T/Z)\sin(k^\parallel\ell)\mathcal{J}_n(qE_0\ell/\omega)$.

Applying Floquet theory, we solve the set of equations~\eqref{firstorder} 
via an analogous expansion~\eqref{channels} into Floquet channels. 
To this end, let us combine the four components $f^{IJ}_{\mathbf{k}s}$
into one vector ${\cal F}_{\mathbf{k}s}$ and write~\eqref{firstorder} 
as 
\begin{align}
i\partial_t
{\cal F}_{\mathbf{k}s}(t)
=
{\cal M}_{\mathbf{k}}(t)
\cdot 
{\cal F}_{\mathbf{k}s}(t)
+
{\cal S}_{\mathbf{k}}(t)
\,,
\end{align}
with $4\times4$ matrices ${\cal M}_{\mathbf{k}}(t)$.
The source terms ${\cal S}_{\mathbf{k}}(t)$ can be eliminated by a 
simple shift of the ${\cal F}_{\mathbf{k}s}$. 
Now we may expand ${\cal M}_{\mathbf{k}}(t)$ and 
${\cal F}_{\mathbf{k}s}(t)$ into Floquet channels as in~\eqref{channels} 
and arrive at 
\begin{align}\label{FloqODE}
i\partial_t{\mathfrak F}_{\mathbf{k}s}(t)
=
{\mathfrak M}_{\mathbf{k}}\cdot{\mathfrak F}_{\mathbf{k}s}(t)
\,,
\end{align}
where ${\mathfrak F}_{\mathbf{k}s}(t)$ is now an infinite dimensional 
vector containing all the Floquet channels $n\in\mathbb Z$. 
In analogy, the matrix ${\mathfrak M}_{\mathbf{k}}$ is infinite 
dimensional as it contains the coupling between all the 
Floquet channels -- but it is independent of time. 
The eigenvectors of the matrix ${\mathfrak M}_{\mathbf{k}}$ correspond 
to the Floquet states whereas the associated eigenvalues 
$\omega\nu_\mathbf{k}$ yield the Floquet exponents $\nu_\mathbf{k}$. 

Translating the above analysis back to our problem suggests the 
following ansatz for the correlators
\begin{align}
f^{01}_{\mathbf{k},s}
&=
\sum_n F^{01}_{\mathbf{k},n}e^{i (n+\nu_\mathbf{k})\omega t}
\label{f01}
\,,
\\
f^{00}_{\mathbf{k},s}+\frac{1}{4}
&=
\sum_n G_{\mathbf{k},n}e^{i (n+\nu_\mathbf{k})\omega t}
=
-f^{11}_{\mathbf{k},s}+\frac{1}{4}
\label{g}
\,,
\\
f^{10}_{\mathbf{k},s}
&=
\sum_n F^{10}_{\mathbf{k},n}e^{i (n+\nu_\mathbf{k})\omega t}
\label{f10}
\,.
\end{align}
Here, we exploited the particle-hole symmetry 
$f^{00}_{\mathbf{k},s}=-f^{11}_{\mathbf{k},s}$ which is a direct consequence of 
equations~\eqref{firstorder}. 

%
%

For the eigenvalue equation 
\begin{align}
\label{eigensystem}
{\mathfrak M}_{\mathbf{k}}\cdot{\mathfrak F}_{\mathbf{k}s}^{\nu_\mathbf{k}}
=
\omega\nu_\mathbf{k}
{\mathfrak F}_{\mathbf{k}s}^{\nu_\mathbf{k}}
\,,
\end{align}
%
it can be shown that the $\nu_\mathbf{k}$ values are real, 
and the solutions always occur in pairs.
Specifically, for each solution with Floquet exponent $\nu_\mathbf{k}$ 
and eigenvector ${\mathfrak F}_{\mathbf{k}s}^{\nu_\mathbf{k}}$
containing the expansion coefficients $F^{01}_{\mathbf{k},n}$, 
$F^{10}_{\mathbf{k},n}$, and $G_{\mathbf{k},n}$, 
there exists a conjugate solution with the opposite Floquet exponent 
$\tilde{\nu}_\mathbf{k}=-{\nu}_\mathbf{k}$.  
Its eigenvector 
$\tilde{\mathfrak F}_{\mathbf{k}s}^{\tilde\nu_\mathbf{k}}$
contains the entries 
\begin{align}
\tilde{F}^{01}_{\mathbf{k},n}=\left({F}^{10}_{\mathbf{k},-n}\right)^*
\,,\;
\tilde{F}^{10}_{\mathbf{k},n}=\left({F}^{01}_{\mathbf{k},-n}\right)^*
\,,\;
\tilde{G}_{\mathbf{k},n}=\left({G}_{\mathbf{k},-n}\right)^*
\,.
\end{align}
%
%
%
Furthermore, from 
$[\langle c^\dagger_{\mu s I}c_{\nu s J}\rangle^\mathrm{corr}]^*=\langle c^\dagger_{\nu s J}c_{\mu s I}\rangle^\mathrm{corr}$, we deduce the 
property $f^{IJ}_{\mathbf{k},s}=(f^{JI}_{\mathbf{k},s})^*$. 
As a result, the Floquet correlators are linear combinations 
of the conjugate solutions
\begin{align}\label{physf01}
f_{\mathbf{k},s}^{01}&=\sum_n \left(F^{01}_{\mathbf{k},n}e^{i(n+\nu_\mathbf{k})\omega t}
+\left(F^{10}_{\mathbf{k},n}\right)^*e^{-i(n+\nu_\mathbf{k})\omega t}\right)\nonumber\\
&=\left(f_{\mathbf{k},s}^{10}\right)^*
\end{align}
and
\begin{align}\label{physf00}
f_{\mathbf{k},s}^{00}+\frac{1}{4}&=\sum_n \left(G_{\mathbf{k},n}e^{i(n+\nu_\mathbf{k})\omega t}
+\left(G_{\mathbf{k},n}\right)^*e^{-i(n+\nu_\mathbf{k})\omega t}\right)\nonumber\\
&=-f_{\mathbf{k},s}^{11}+\frac{1}{4}\,.
\end{align}
A remaining issue concerns the proper normalization of the correlators. 
Since we applied the mode expansion to correlation functions instead 
of a state vector of the system, the eigenvectors 
${\mathfrak F}_{\mathbf{k}s}^{\nu_\mathbf{k}}$ 
are not normalized to unity. 
Therefore, an additional condition is required to specify the 
appropriate normalization of the correlators.

In the case of the mean-field background~(\ref{disorder}), we observe 
that the evolution equations~(\ref{firstorder}) leave the quantity 
$f_{\mathbf{k},s}^{01}f_{\mathbf{k},s}^{10}
+(f_{\mathbf{k},s}^{00}+1/4)^2=1/16$ invariant. 
From this condition, we deduce that in leading order of the hierarchy, 
the quasi-particle modes remain bounded regardless of the form of 
the applied electrical field.
Furthermore, we can derive the normalization condition 
\begin{align}\label{norm}
 \sum_n\big[|F_{\mathbf{k},n}^{01}|^2+|F_{\mathbf{k},n}^{10}|^2+2|G_{\mathbf{k},n}^{00}|^2\big]=\frac{1}{16}
 \,,
\end{align}
where we have employed the Fourier expansions~(\ref{physf01}) 
and~(\ref{physf00}). 
It is important to note that this condition explicitly differs from 
the usual vector norm of 
${\mathfrak F}_{\mathbf{k}s}^{\nu_\mathbf{k}}$. 

%

\subsection{Floquet exponents}

The eigenvalue equation~(\ref{eigensystem}) has solutions for both 
integer and non-integer values of $\nu_\mathbf{k}$. 
Furthermore, given any solution $\nu_\mathbf{k}$, one can construct 
other solutions with $\nu_\mathbf{k}\pm1$ by a simple transformation
of the eigenvectors ${\mathfrak F}_{\mathbf{k}s}^{\nu_\mathbf{k}}$, 
see Appendix \ref{app}. 
If we now focus on the non-integer Floquet exponents near zero, i.e.,  $0<\nu_\mathbf{k}\ll1$, they are related to the characteristic time scale 
$\tau_\mathbf{k} \sim1/(\nu_\mathbf{k}\omega)$ of the growth of the 
correlation functions, for example near a resonant frequency $\omega=U/n$. 

For small hopping, specifically $T/\omega\ll1$ and $T/U\ll1$, 
it is possible to determine these Floquet exponents from the 
eigenvalue equation (\ref{eigensystem}).
After some algebra (see Appendix \ref{app}), we obtain the following expression \cite{W89}:
\begin{widetext}
\begin{align}\label{det}
\sin^2(\pi \nu_\mathbf{k})
=
\sin^2\left(\frac{\pi U}{\omega}\right)
+\sin\left[\frac{2\pi U}{\omega}\right]\frac{\pi}{2\omega}
\left(
\frac{(T_{\mathbf{k}}^\perp)^2}{U}
+\frac{2T_{\mathbf{k}}^\perp T_{\mathbf{k},0}^\|}{U}
+\sum_n |T_{\mathbf{k},n}^\||^2\frac{U}{U^2-\omega^2 n^2}
\right)+\mathcal{O}(T^4)
\,.
\end{align}
\end{widetext}
Note that the apparent singularity from the denominator 
$U^2-\omega^2 n^2$ at resonance $\omega=U/n$ is compensated by the 
pre-factor $\sin(2\pi U/\omega)$ which also vanishes at $\omega=U/n$.

When the driving frequency approaches the resonance condition
$\omega=U/n$, the above relation yields a simple scaling law 
for small electric fields $qE_0\ell\ll U$
\begin{align}
\nu_\mathbf{k}\approx 
\frac{\sqrt{2} nT}{ZU}
\frac{1}{ 2^n n!}
\left(\frac{qE_0\ell n}{U}\right)^n
\begin{cases}
\cos(k^\parallel\ell) &\text{ for $n$ even} \\
\sin(k^\parallel\ell) &\text{ for $n$ odd}\,,
\end{cases}
\end{align}
where we have inserted the results for the square lattice as a specific
example. 

More generally, near the $n$-th resonance, the Floquet exponent scales 
with $E_0^n$.
This is consistent with the intuitive picture that absorbing $n$ 
photons is necessary to overcome the energy gap $U$ of the system. 
Since the process of multi-photon absorption (for weak fields $E_0$) 
should occur on a timescale that increases with the number of 
photons involved, we expect a slower increase around the 
higher-order resonances. 
Indeed, the characteristic timescale in which the correlations build up 
scales with powers of the electrical field, specifically 
$\tau_\mathbf{k}\sim (\omega/E_0)^n(Z/T)$.

\section{Higher harmonics}

After having developed the Floquet analysis of the doublon-holon 
correlators in the driven Fermi-Hubbard mode, we may now turn to 
specific applications.

For example, one might envision switching on the pump laser suddenly,
which corresponds to a quantum quench.
After preparing the Hubbard system initially in the ground state 
(with respect to the doublon-holon quasi-particle excitations, 
not the spin structure), 
the system is driven into an excited state far from equilibrium
by this sudden switching. 
Within our level of approximation, this initial state is then translated 
into a superposition of eigenstates of the Floquet system~(\ref{eigensystem})
for each mode $\mathbf{k},s$. 
The various eigenvalues $\nu_\mathbf{k}$ occurring in this superposition
would then determine the individual oscillation or initial growth of 
these Floquet states. 

In the following, we consider a slow switching of the pump laser instead 
-- which is probably closer to a typical experimental situation.
Again starting in the ground state, the system is stationary for $E_0=0$, 
and thus the left-hand side of~(\ref{firstorder}) will vanish. 
As a result, the system will be in a state with zero eigenvalue 
$\nu_\mathbf{k}=0$.
If the pump laser is now turned on adiabatically -- as described by 
a matrix ${\mathfrak M}_{\mathbf{k}}(t)$ with a very slow time-dependence 
-- the system will remain in the Floquet state corresponding to 
$\nu_\mathbf{k}=0$. 
This, together with the normalization condition~(\ref{norm}), 
allows us to evaluate the correlators 
$F_{\mathbf{k},n}^{01}$, $F_{\mathbf{k},n}^{10}$, and $G_{\mathbf{k},n}$.

To relate our findings to observable quantities, we consider the 
particle current between neighboring lattice sites $\mu$ and $\nu$
(for which $T_{\mu\nu}=T$) 
\begin{align}
J=i\sum_s
\langle \hat{c}^\dagger_{\mu,s}\hat{c}_{\nu,s}\rangle+\rm h.c.
\end{align}
In general, this current could also have a component perpendicular 
to the pump field $\mathbf{E}_0$, but for our example of the 
square lattice, it is purely parallel and reads 
\begin{align}\label{current}
J=-4\int_\mathbf{k}\sin(k^\parallel\ell)
(f^{01}_{\mathbf{k},s}+f^{10}_{\mathbf{k},s})
\,.
\end{align}
Due to the anti-symmetry of the integrand in~(\ref{current}) with respect 
to $k^\parallel$ and the symmetry properties of the $T_{\mathbf{k},n}^\|$, 
only odd multiples of higher harmonics occur, and we have 
$J(t)=\sum_n J_n\sin(n\omega t)$, where $n=1,3,5,...$.

Close to a resonance at $\omega n=U$, the current is dominated by a 
single correlator
\begin{align}
F_{\mathbf{k},n}^{10}
\approx
-\frac{U T_{\mathbf{k},n}^\|}
{2\sqrt{T_\mathbf{k}^4+
4U^2\left|T_{\mathbf{k},n}^\|\right|^2}}
\,,
\end{align}
while all other correlators are suppressed by at least a factor of $T/U$. 
Evaluating the integration over the Brillouin zone for small electrical 
fields ($qE_0\ell/U\ll1$), we obtain the resonant current as
\begin{align}
J(t)
\approx 
1.4 
\sqrt{\frac{U}{2T}\left(\frac{nqE_0\ell}{U}\right)^n\frac{1}{2^{n}n!}}
\,\sin(U t) \,. 
\end{align}
The non-polynomial dependence of the current on the involved quantities 
$U$, $T$ and $E_0$ indicates that it should be non-trivial to obtain 
this result via some sort of perturbation theory.
Time-dependent perturbation theory could be used to derive the initial 
growth of correlations after suddenly turning on the pump laser 
(as described above), but here we are considering a steady state. 

For the fundamental resonance $\omega=U$, the current scales with 
$\sqrt{qE_0\ell/U}$. 
This current would then emit electromagnetic radiation with the same 
frequency $\omega=U$ as the incident light. 
In order to emit higher frequencies, one could use the 
third harmonic $\omega=U/3$ where the current 
is suppressed by an additional factor of $qE_0\ell/U$. 

%
%

\section{conclusions and outlook}

By combining Floquet theory with the method of the hierarchy of 
correlations, we study the non-equilibrium dynamics of the driven 
Fermi-Hubbard model under the influence of a harmonically 
oscillating electric field, which serves as a model for the pump 
laser.  
We derive the associated Floquet exponents, which determine the time 
scales of the evolution, such as the initial growth of the correlations
after suddenly switching on the pump laser. 

After developing the basic theory, we study higher-harmonic generation
in a steady state, which can be reached by gradually turning on the 
pump laser.
To this end, we calculate the time-dependent current which can be 
obtained from the correlations (or coherences) between neighboring sites.
The ability to calculate such a quantity involving operators at two 
lattice sites highlights one of the advantages of the method of the 
hierarchy of correlations in comparison to other approaches which 
effectively map the system onto a single lattice site 
(such as dynamical mean-field theory DMFT \cite{A14,G96,S19,P18,Q18,J08,D23}). 

As an outlook, our results can be used as an ingredient to study the 
energy balance of the steady Floquet states, where the energy absorption 
from the pump laser is compensated by the emission (including the 
higher-harmonics).
Going from quasi-stationary Floquet states to more general 
non-stationary states (i.e., superpositions of Floquet states),
the formalism developed above can als be used to investigate 
pre-thermalization \cite{H17,R20,K11,K08,H13,Q14} and other phenomena. 

\acknowledgments 

Funded by the Deutsche Forschungsgemeinschaft 
(DFG, German Research Foundation) -- Project-ID 278162697-- SFB 1242. 

 
\appendix

\section{Floquet eigensystem}\label{app}
 
Utilizing the Floquet expansions (\ref{f01}) - (\ref{f10}) 
 for the correlators, the system of algebraic equations adopts the following structure%
\begin{align}
&-\omega (n+\nu_\mathbf{k})G_{\mathbf{k},n}(\nu_\mathbf{k})-\frac{T^\perp_{\mathbf{k}}}{2}[F^{10}_{\mathbf{k},n}(\nu_\mathbf{k})-F^{01}_{\mathbf{k},n}(\nu_\mathbf{k})]\nonumber\\
&=\frac{1}{2}\sum_m T^\parallel_{\mathbf{k},m} [F^{10}_{\mathbf{k},n+m}(\nu_\mathbf{k})-F^{01}_{\mathbf{k},n+m}(\nu_\mathbf{k})]\label{FH1}\\
&\left[-\omega(n+\nu_\mathbf{k})-U\right]F^{01}_{\mathbf{k},n}(\nu_\mathbf{k})+T^\perp_{\mathbf{k}}G_{\mathbf{k},n}(\nu_\mathbf{k})\nonumber\\
&=-\sum_m T^\parallel_{\mathbf{k},m} G_{\mathbf{k},n+m}(\nu_\mathbf{k})\label{FH2}\\
&\left[-\omega(n+\nu_\mathbf{k})+U\right]F^{10}_{\mathbf{k},n}(\nu_\mathbf{k})-T^\perp_{\mathbf{k}}G^{00}_{\mathbf{k},n}(\nu_\mathbf{k})\nonumber\\
&=\sum_m T^\parallel_{\mathbf{k},m} G_{\mathbf{k},n+m}(\nu_\mathbf{k})\label{FH3}
\end{align}
Here, we have explicitly denoted that the coefficients belong to the eigenvector with eigenvalue $\nu_\mathbf{k}$.

If our objective is to derive the eigenvector for the eigenvalue $\nu_\mathbf{k}+l$, where $l$ is a natural number, we can achieve this by performing an index shift $n\rightarrow n-l$, resulting in the following correspondences
\begin{align}
F^{01}_{\mathbf{k},n}(\nu_\mathbf{k}+l)&=F^{01}_{\mathbf{k},n+l}(\nu_\mathbf{k})\\
F^{10}_{\mathbf{k},n}(\nu_\mathbf{k}+l)&=F^{10}_{\mathbf{k},n+l}(\nu_\mathbf{k})\\
G_{\mathbf{k},n}(\nu_\mathbf{k}+l)&=G_{\mathbf{k},n+l}(\nu_\mathbf{k})\,.
\end{align}
This naturally signifies that the system's behavior is entirely defined by its values within the Floquet-Brillouin zone.

A solution exists for the system of homogeneous equations (\ref{FH1})-(\ref{FH3}) if the determinant of $\mathfrak{M} + \omega \nu_\mathbf{k} \mathbb{1}$ zero (cf. equation \ref{FloqODE}).
In instances where the Floquet exponent is not an integer, it is possible to normalize the rows of this infinite-dimensional matrix in such a way that the diagonal elements become unity.
After employing the particle-hole symmetry, we subsequently obtain a matrix that follows the pattern

\begin{align}
\mathfrak{N}=\mathbb{1}&+
\begin{pmatrix}
                          ...&... &... &... & ...\\
                          ...&D_{n-1} & 0 & 0 &...\\
                          ...&0 & D_{n} & 0 &...\\
                          ...&0 & 0 & D_{n+1} &...\\
                          ...&... &... &... & ...\\                          
                          \end{pmatrix}
                          \nonumber\\
&+\begin{pmatrix}
                          ...&... &... &... & ...\\
                          ...&C_{n-1,0} & C_{n-1,+1} & C_{n-1,+2} &...\\
                          ...&C_{n,-1} & C_{n,0} & C_{n,+1} &...\\
                          ...&C_{n+1,-2} & C_{n+1,-1} & C_{n+1,0} &...\\
                          ...&... &... &... & ...\\                          
                          \end{pmatrix}
\end{align}
with the $3\times 3$ matrices
\begin{align}
C_{n,m}=
\begin{pmatrix}
0 &-\frac{T^\parallel_{\mathbf{k},m}}{\omega(n+\nu_\mathbf{k})+U} & 0\\
-\frac{T^\parallel_{\mathbf{k},m}}{2\omega(n+\nu_\mathbf{k})} & 0 & \frac{T^\parallel_{\mathbf{k},m}}{2\omega(n+\nu_\mathbf{k})}\\
0 & \frac{T^\parallel_{\mathbf{k},m}}{\omega(n+\nu_\mathbf{k})-U} & 0
\end{pmatrix}
\end{align}
and
\begin{align}
D_{n}=
\begin{pmatrix}
0 &-\frac{T^\perp_{\mathbf{k}}}{\omega(n+\nu_\mathbf{k})+U} & 0\\
-\frac{T^\perp_{\mathbf{k}}}{2\omega(n+\nu_\mathbf{k})} & 0 & \frac{T^\perp_{\mathbf{k}}}{2\omega(n+\nu_\mathbf{k})}\\
0 & \frac{T^\perp_{\mathbf{k}}}{\omega(n+\nu_\mathbf{k})-U} & 0
\end{pmatrix}
\end{align}
The determinant can be computed perturbatively in the limit of strong interactions and we find

\begin{align}
0\overset{!}{=}&\mathrm{Det}(\mathfrak{N})=1-\frac{1}{2}\bigg[\sum_n \mathrm{Tr}(D_n^2)+2\sum_n \mathrm{Tr}(D_n C_{n,0})\nonumber\\
&+\sum_{n,m}\mathrm{Tr}(C_{n,m}C_{n+m,-m})\bigg]+\mathcal{O}\left(T^4\right)\,.
\end{align}
Upon the evaluation of traces and summations, we ultimately reach equation (\ref{det}).

\end{document}